\documentstyle[12pt,aaspp4]{article}
\begin{document}

\title{Submillimeter observations of IC 10: the dust properties and
neutral carbon content of a low metallicity starburst}

\author{Alberto D. Bolatto, James M. Jackson} \affil{Institute for
Astrophysical Research, Department of Astronomy, Boston University,
Boston MA 02215} \authoraddr{Boston University Dept. of Astronomy\\725
Commonwealth Ave.\\ Boston, MA 02215} \author{Christine D. Wilson}
\affil{Department of Astronomy, McMaster University, Ontario, Canada}
\and \author{Gerald Moriarty-Schieven} \affil{Joint Astronomy Centre,
Hilo, Hawaii}

\begin{abstract}
We present submillimeter observations of the Local Group, metal-poor,
irregular dwarf galaxy IC~10, directly relevant to the interaction
between interstellar medium and star formation activity in primeval
galaxies. Using the JCMT, we have observed the fine structure neutral
carbon transition $^3$P$_1\rightarrow^3$P$_0$ at 492 GHz and the
rotational $J=3\rightarrow2$ transition of $^{12}$CO and $^{13}$CO in
the most massive giant molecular cloud complex in this galaxy,
IC~10-SE.  We find that, although the $\rm I_{[CII]}/I_{CO}$ ratio for
this object is a factor of 4 larger than the typical Milky Way value,
its [C~I] to CO intensity ratio $\rm I_{[CI]}/I_{CO}\simeq18\pm2$ (in
units of erg s$^{-1}$ cm$^{-2}$ sr$^{-1}$) is similar (only about 50\%
larger) to that of the Milky Way.  Modelling of the behaviour of the
[C~II]/CO and [C~I]/CO intensity ratios with metallicity indicates
that, if C$^+$ and C$^0$ are chiefly produced by UV photodissociation
in the PDR, both ratios should increase sharply with decreasing
metallicity (and consequently diminished UV shielding; Bolatto,
Jackson, \& Ingalls 1999\markcite{BJI99}). These data then suggest a
different origin for an important fraction of C$^0$ in these clouds,
unrelated to photodissociation.

We have also mapped the 850 $\mu$m continuum in this region using
SCUBA. Employing these data in conjunction with KAO and IRAM
measurements we find that the 100 $\mu$m to 1300 $\mu$m continuum
emission corresponds to a graybody with an extremely low emissivity
exponent, $\beta\sim0.5$.  We conclude that this low exponent is most
likely due to the destruction of small dust grains, brought about by
the increased penetration of UV radiation in the low metallicity
ISM. If a low emissivity exponent in the submillimeter is a general
property of metal-poor systems then the interpretation of millimeter
and submillimeter surveys of high-z galaxies should be revised.
\end{abstract}


\section{Introduction}
\label{intro}

IC~10, an irregular, low metallicity dwarf galaxy (IBm or ``Magellanic
Cloud'' type), is one of the most active star-forming galaxies in the
Local Group.  Its galaxy-wide massive star surface density is
comparable to that of the starburst regions of M~33 (Massey,
Armandroff, \& Conti 1992\markcite{MA92}). Because IC~10 is a nearby
system that in many ways resembles primeval galaxies (morphologically
irregular, metal-poor, and very actively forming stars) we have
targeted it for a study of the interstellar medium (ISM) at
submillimeter wavelengths that has direct relevance to star formation
at high redshifts.  This galaxy, at a distance of 0.82 Mpc (Wilson et
al. 1996\markcite{WI96}), is comparable in mass and size to the Small
Magellanic Cloud. Its metallicity ($12+\log{\rm(O/H)}=8.17$,
$Z/Z_{\odot}\simeq\slantfrac{1}{6}$ Lequeux et
al. 1979\markcite{LE79}) falls between that of the Large and the Small
Magellanic Clouds.

IC~10 is an ideal laboratory to study the interaction between a
metal-poor ISM and strong radiation fields.  Metallicity greatly
influences the composition and structure of the ISM, especially when
intense UV fields are present. Photodissociation regions (PDRs) are
those portions of the interstellar medium where UV radiation dominates
the physical and chemical processes. Because low metallicity PDRs have
fewer heavy elements their gas phase chemistry is altered, changing
the equilibrium abundances of the molecular species (e.g., van
Dishoeck \& Black 1988\markcite{VDB88}; Lequeux et
al. 1994\markcite{LE94}).  The deficiency of heavy elements also
lowers the dust-to-gas ratio.  This in turn allows the UV radiation to
penetrate more deeply into the molecular material, thereby causing
widespread photodissociation and photoionization.  The interaction
between the starburst and the ISM in a low metallicity system is thus
maximized, as the UV photons permeate the ISM relatively unimpeded.
Molecular hydrogen is largely unaffected by this enhanced UV radiation
because of the strong H$_2$ self-shielding and the mutual shielding
between coincident lines of H and H$_2$ (Abgrall et
al. 1992\markcite{AB92}), but CO is strongly photodissociated.
Ionized carbon [C~II] observations in IC~10 suggest the presence of
vast quantities of molecular hydrogen not traced by CO clouds (Madden
et al. 1997\markcite{MA97}).

IC~10 is undergoing a burst of star formation, with the center of
activity in its south-east portion.  Nonthermal emission, probably
arising from a collection of supernova remnants (Yang \& Skillman
1993\markcite{YS93}), H$_2$O megamaser emission (Henkel, Wouterloot,
\& Bally 1986\markcite{HE86}; Baan \& Haschick 1994\markcite{BH94}),
one of the highest surface densities of Wolf-Rayet stars in the Local
Group (Massey et al. 1992\markcite{MA92}) and one of the largest FIR
luminosities among dwarf galaxies (Melisse \& Israel
1994\markcite{MI94}) all suggest current vigorous star formation
activity. The ratio of WC types (He burning Wolf-Rayet stars) to WN
types (CNO cycle Wolf-Rayet stars) is also unusually high. This has
been interpreted as an initial mass function skewed towards higher
stellar masses (Massey \& Armandroff 1995\markcite{MA95}).

Neutral hydrogen maps of IC~10 show a complex structure characterized
by many peaks and holes, an extended low brightness halo, and a
complex velocity field (Shostak 1974\markcite{SH74}; Shostak \&
Skillman 1989 \markcite{SS89}; Wilcots \& Miller
1998\markcite{WM98}). The brightest H~I peak is located in the
southern part of the galaxy and it is associated with a cluster of
H~II regions (Hodge \& Lee 1990\markcite{HL90}) and a molecular cloud
complex named IC~10-SE (Becker 1990\markcite{BE90}).  The center of
star formation activity in IC~10 is located in close proximity to this
H~I column density maximum.

The CO emission from IC~10 is relatively weak, as is usual in dwarf
irregular galaxies (e.g., Tacconi \& Young 1987\markcite{TY87}), many
of which appear to be deficient in molecular gas although quite
actively forming stars.  CO was first detected by Henkel et al.
(1986)\markcite{HE86}. The molecular cloud complexes have been mapped
using single dish observations by Ohta, Sasaki, \& Sait\=o
(1988)\markcite{OH88} and Becker (1990)\markcite{BE90}, and
interferometrically by Wilson \& Reid (1991)\markcite{WR91}, Ohta et
al. (1992)\markcite{OH92} and Wilson (1995)\markcite{WI95}. The CO
intensity to H$_2$ column density conversion factor $X_{\rm CO}$
strongly depends on the metallicity of the clouds (Wilson
1995\markcite{WI95}; Arimoto, Sofue, \& Tsujimoto 1996\markcite{AR96};
Israel 1997\markcite{IS97}). This dependency explains why some dwarf
irregular galaxies appear to be forming stars much more efficiently
than the Milky Way; much of their molecular gas is not well traced by
CO.

In this paper we present observations of the [C~I] ($J=1\rightarrow0$)
and CO ($J=3\rightarrow2$), two of the most important submillimeter
cooling lines of the ISM, together with the dust continuum of the
molecular cloud complex IC~10-SE. Neutral carbon is the byproduct of
the photodissociation of CO, created when UV photons impinge on the
surfaces of molecular clouds. Thus the [C~I]
$^3$P$_1\rightarrow^3$P$_0$ fine structure transition is thought to
trace the photodissociation region (PDR), the transition zone at low
extinction ($A_v\sim1$---3) where UV radiation dissociates the skin of
CO clumps.  The CO ($J=3\rightarrow2$) rotational transition typically
traces moderately warm ($T\sim30$ K) and dense ($n\sim4\times10^4$
cm$^{-3}$) molecular gas, although determining a precise $T$ and $n$
requires detailed modelling of several spectral lines.

\section{Observations and results}
\subsection{Spectroscopic observations}
We observed the [C~I] fine structure line at 492.1607 GHz, the
$^{12}$CO ($J=3\rightarrow2$) rotational transition at 345.7960 GHz
and the $^{13}$CO ($J=3\rightarrow2$) rotational line at 330.5879 GHz
toward IC~10-SE using the 15 meter James Clerk Maxwell Telescope
(JCMT) at Mauna Kea, Hawaii. These observations were performed in the
remote observing mode, during 8 half shifts from 18 to 28 July
1997. Pointing and focus were checked at the beginning of each 4 hour
long observing segment.  Standard flux calibrators were observed
before and after the IC~10 observations, and their intensities were
found to be within 30\% ($3\sigma$) of their accepted values.
Calibration was performed every 20 minutes using the standard chopper
wheel method with warm (ambient temperature) and cold loads.

The [C~I] line was observed using the RxC2 SIS receiver, with
$\eta_{mb}=0.52$ and HPBW=10.8''.  The $\tau_{225}$ zenith opacity was
in the range 0.04 to 0.05, resulting in system temperatures
$T_{sys}=2500 - 3500$ K.  When the weather did not allow [C~I]
observations we switched to our backup program, observing CO
($J=3\rightarrow2$) with the RxB3 SIS receiver, $\eta_{mb}=0.64$ and
HPBW=13.2''.  The $\tau_{225}$ opacity at this time was in the range
0.12 to 0.16 with corresponding system temperatures $T_{sys}=800 -
1500$ K for $^{12}$CO and $T_{sys}\simeq1400$ K for $^{13}$CO.  The
back-end was the Digital Autocorrelation Spectrometer (DAS).  The
[C~I] observations were performed in the DAS 500 MHz bandwidth mode,
resulting in a spectral resolution of 374 kHz (0.23 km s$^{-1}$). The
CO observations were accomplished using the DAS 760 MHz bandwidth mode
with a resolution of 750 kHz (0.65 km s$^{-1}$).

We observed [C~I] towards 6 positions in the molecular cloud complex
IC~10-SE. Three of them correspond to the three clumps identified by
Wilson \& Reid (1991) using CO ($J=1\rightarrow0$) interferometric
observations. The other three positions were arranged to obtain a
slice across the complex, along an axis roughly oriented towards the
nearest cluster of H~II regions.  The [C~I] observations of clumps
MC1, MC2 and MC3 on IC~10-SE (Wilson, \& Reid 1991\markcite{WR91})
were performed in position switching mode, switching 30' away in
azimuth. The rest of the observations were performed in beam switching
mode, chopping with the secondary mirror 180'' away at a frequency of
1 Hz. This technique resulted in better baselines with no obvious
problems of emission in the reference position.

\begin{figure}
\plotone{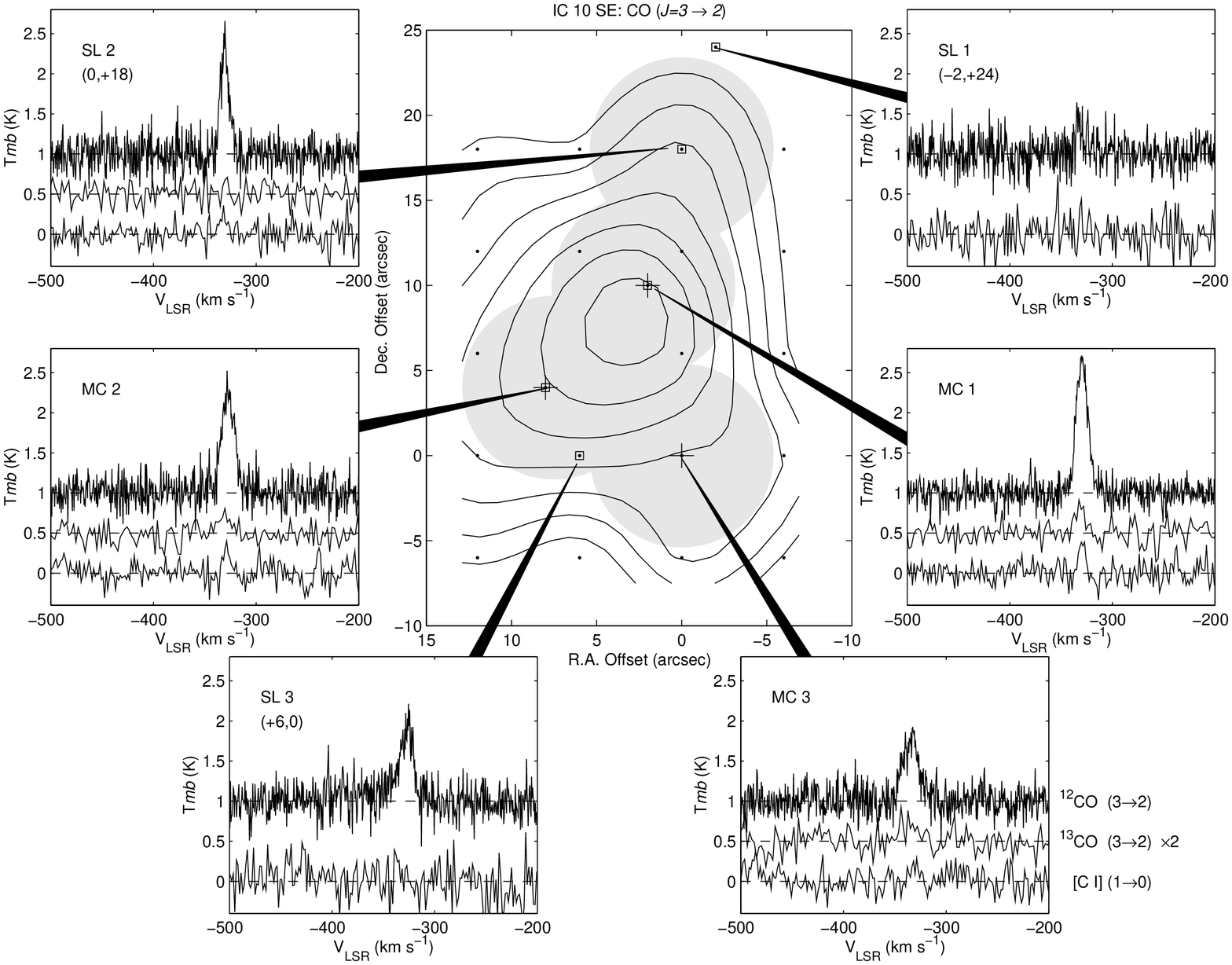} \figcaption[fig1.eps]{[C~I] and CO
($J=3\rightarrow2$) in IC~10-SE. The contour map shows the CO
integrated intensity. The contour levels are 8 to 23 in steps of 2.5 K
km s$^{-1}$, and the map offsets are with respect to MC3 (c.f., Table
\protect\ref{tabci}).  The dots show the actual CO pointings that
constitute the map. The crosses are placed at the positions of clumps
identified by Wilson \& Reid (1991)\protect\markcite{WR91}.  The
squares show the placement of our slice observations, while the gray
circles illustrate the size of the JCMT [C~I] beam. Our six [C~I]
spectra (lower) are shown here together with the $^{12}$CO (upper) and
$^{13}$CO ($J=3\rightarrow2$) (middle) obtained towards the same
positions. For display purposes, the $^{13}$CO spectra are scaled by a
factor of 2 and displaced 0.5 K, while the $^{12}$CO spectral are
displaced by 1.0 K.\label{spectra}}
\end{figure}

First order baselines were removed from each spectrum and the
individual 5 minute integrations were coadded.  After averaging, the
[C~I] spectra were Hanning smoothed to 3 MHz ($\sim 1.85$ km s$^{-1}$)
.  The results are shown in Figure \ref{spectra} and compiled in
Tables \ref{tabci} and \ref{tabco}.

\subsection{Continuum observations}
The Submillimetre Common User Bolometer Array (SCUBA) 850 \micron\ and
1350 \micron\ observations were performed in service mode during the
nights of 27 September, 10 November and 23 December 1997. SCUBA's 850
\micron\ and 1350 \micron\ beam sizes are respectively 14'' and 18''
(Holland et al. 1999 \markcite{HO99}).  One 64-position jiggle map was
obtained at 850 \micron\ (Fig. \ref{scuba}), by imaging a field
$\sim2.3$ arcmin in diameter centered NE of the IC~10-SE molecular
cloud complex and containing it, as well as several nearby H~II
regions (Hodge \& Lee 1990\markcite{HL90}). The observations were
performed in three different sessions, with a typical zenith sky
transmission at 850 \micron\ of 70\% ($\tau_{850}\simeq0.4$).  We used
a chopper throw of 120'' in azimuth, and the total accumulated
integration time was 128 minutes.  Pointing was checked before each
observing session and found to be within 2''.  Calibration was
performed using Uranus and Mars.

Simultaneous 450 \micron\ data were acquired using SCUBA's
short-wavelength array, but due to the poor weather ($\tau_{450}>2$)
these observations did not yield any useful data.  The 1350 \micron\
observations were performed using one of SCUBA's photometric pixels,
acquiring one point at a time in a five point map centered on IC~10-SE
and chopping 60'', with an integration of 3 minutes per point.  Only
the center position ($\alpha_{1950}=00^h17^m44\fs5$,
$\delta_{1950}=59^\circ00'22''$) yielded a significant detection with
a flux density of $55\pm5$ mJy.

\section{Discussion}
\subsection{Comparison with previous CO observations}
\label{comparison}

In this section we compare our CO observations with existing
single-dish and interferometric data of IC~10-SE. With this comparison
we aim to: 1) refine our determinations of line ratios by improving
the alignment of the maps and 2) discuss possible excitation gradients
in the source.

\begin{figure}
\plotone{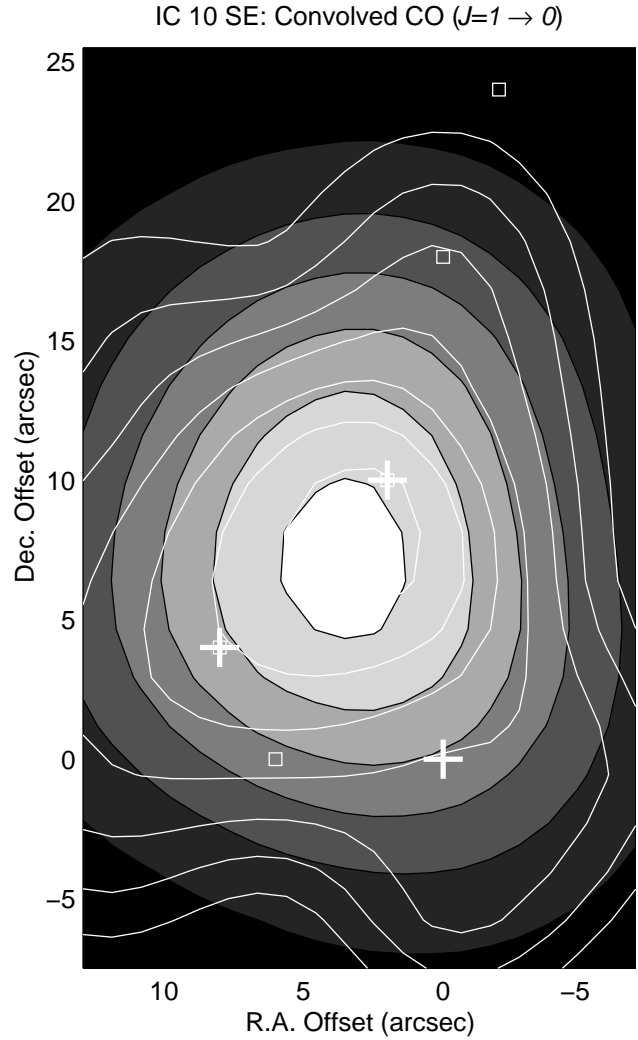} \figcaption[fig2.eps]{CO ($J=3\rightarrow2$)
contour map (white) overlayed on the OVRO interferometric CO
($J=1\rightarrow0$) (Wilson \& Reid 1991 \protect\markcite{WR91})
convolved to the JCMT CO ($J=3\rightarrow2$) beam (gray scale
contours). The OVRO data has been displaced 2'' to the north from its
nominal coordinates.  The OVRO contour levels are 2 to 12 in steps of
2 K km s$^{-1}$.\label{ovroco}}
\end{figure}

Figure \ref{ovroco} shows a comparison of our observations with Wilson
\& Reid's (1991) \markcite{WR91} interferometric $^{12}$CO
($J=1\rightarrow0$) measurements convolved to the JCMT's beam.  The
agreement is extremely good after displacing the OVRO map 2'' to the
north of its nominal coordinates. Such an offset can be caused by
errors in the JCMT pointing model, which has a typical RMS of 1.5'',
or poor phase calibration in the interferometric data which can
introduce displacements of order half a synthesized beam. Although
either telescope can be the source of this discrepancy, in this work
we have chosen to move the OVRO map for convenience.  To compute the
$\rm I_{[C I]}/I_{CO}$ ratios discussed in \S\ref{ratios} we have used
the displaced OVRO data.

\begin{figure}
\plotone{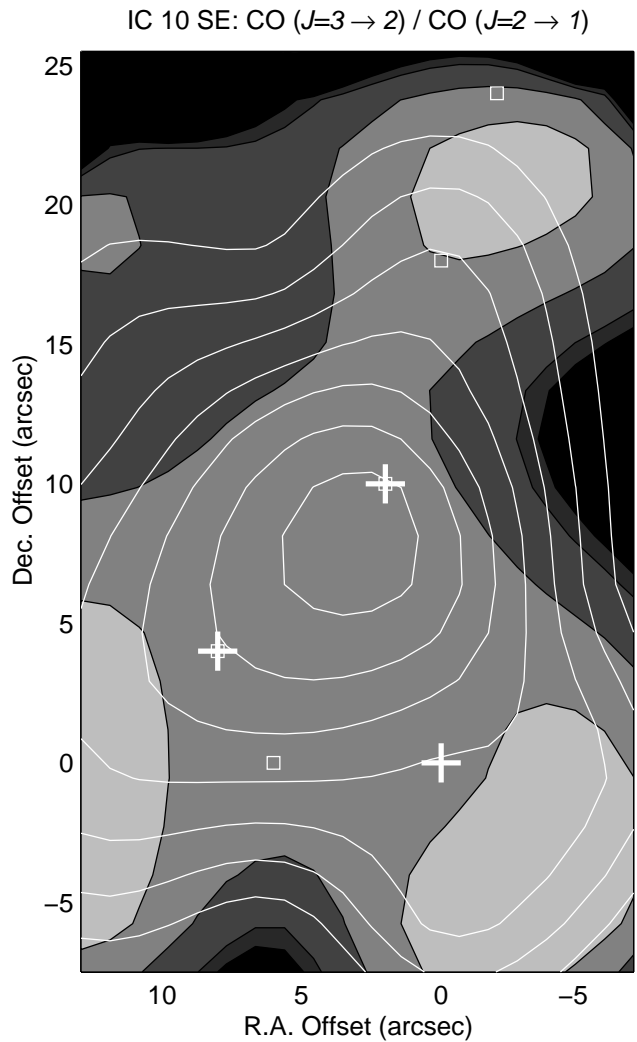} \figcaption[fig3.eps]{Planck-corrected peak antenna
temperature ratio map for the CO ($J=3\rightarrow2$) (this paper) and
CO ($J=2\rightarrow1$) (Becker 1990\protect\markcite{BE90})
transitions (grayscale) overlayed on the CO ($J=3\rightarrow2$)
integrated intensity map (white contours). Both data sets have very
similar spatial resolution. The gray scale contours are 0.8 to 1.1 in
steps of 0.1. This map shows that, within errors, the
$J=3\rightarrow2$ and $J=2\rightarrow1$ transitions are probably
optically thick and thermalized, and they trace the same gas
throughout the complex.\label{texrat}}
\end{figure}

Although Petitpas \& Wilson (1998)\markcite{PW98} studied the CO
excitation in this complex for the central position, our new data
reveal the spatial distribution.  Figure \ref{texrat} shows the ratio
of Planck-corrected peak antenna temperature (i.e., the temperature of
a blackbody that would emit as much power as is detected by the
antenna) for the $J=3\rightarrow2$ and $J=2\rightarrow1$ transitions
(Becker 1990\markcite{BE90}; IRAM 30 meter telescope), computed as,

\begin{equation}
T_{A}^{Planck} = \frac{T^*}{\ln\left(\frac{T^*}{T_A+J_\nu(T_{\rm
CMBR})}+1\right)}
\end{equation}

\noindent where $T^*=h\nu/k$ is the energy of the transition expressed
in temperature units, $T_A$ is the measured Rayleigh-Jeans antenna
temperature, and $J_\nu(T_{\rm CMBR})={T^*}/({\exp(T^*/T_{\rm CMBR})
-1})$ is the Cosmic Microwave Background contribution at the frequency
of interest. Because both sets of observations have very similar beam
sizes no convolution is necessary (note that the IRAM map is not
Nyquist-sampled, therefore we are losing some spatial information).
The Planck-corrected antenna temperature ratio is
$T^{Planck}_{3\rightarrow2}/T^{Planck}_{2\rightarrow1}\approx1$
($T_{mb (3\rightarrow2)}/T_{mb (2\rightarrow1)}\sim$0.55---0.7)
throughout IC~10-SE, consistent with thermalized gas (i.e., gas with
kinetic temperature equal to the excitation temperature,
T$_{ex}=$T$_{kin}$) at a temperature higher than the excitation energy
of the $J=2$ level (i.e., T$_{kin}>15$ K). Thermalization is most
easily attained if the gas is optically thick.

The $^{12}$CO/$^{13}$CO ratio of the integrated line intensities
listed in Table \ref{tab13co} also suggests optically thick $^{12}$CO
gas at the positions of the three main molecular clumps (MC1, MC2 and
MC3).  Although there are no estimates of the $^{12}$C/$^{13}$C
abundance ratio in IC~10, the measured $^{12}$CO/$^{13}$CO line ratio
is $\approx10$ for each of the three clumps, much smaller than the
typical $^{12}$C/$^{13}$C$\approx50$ isotopic ratio in Galactic or LMC
molecular clouds (e.g., Langer \& Penzias 1990\markcite{LP90};
Johansson et al. 1994\markcite{JO94}).  At the position SL2, however,
no $^{13}$CO is detected in spite of the strong $^{12}$CO
emission. This may be caused by a smaller CO optical depth at the edge
of the complex, which is consistent with the lack of strong emission
at the position of SL2 in the CO ($J=1\rightarrow0$) interferometric
map (Wilson \& Reid 1991\markcite{WR91}).

Table \ref{tabco} shows that the linewidths of the $^{12}$CO spectra
are systematically larger than those of [C~I] by a factor of $\sim 2$
at the positions of the main molecular clumps, while the [C~I]/CO
linewidth ratio decreases to $\sim 1$ for SL2, at the edge of the
complex.  The broadening of spectral lines in the ISM is usually
attributed to either optical depth or Doppler (turbulent) effects such
those in macroturbulent models (e.g., Wolfire, Hollenbach \& Tielens
1993\markcite{WHT93}).  The linewidth ratios are thus consistent with
an increasing optical depth towards the center of the complex. In
particular, the linewidth ratio of $\sim1$ towards SL2 is consistent
with the lower CO optical depth deduced from the $^{12}$CO/$^{13}$CO
line ratio.  If the line profiles are instead caused by
macroturbulence the linewidths will be dominated by the clump-to-clump
velocity dispersion.  A [C~I] linewidth systematically smaller than
its CO counterpart can only be understood if the [C~I] emission arises
solely from a fraction of the clumps featuring a smaller velocity
dispersion. This, in turn, would imply that the [C~I] emitting clumps
are localized within a distinct region inside our beam. The
interpretation of the observed [C~I] linewidths as arising from
macroturbulence requires, however, optically thick [C~I] emission from
individual clumps which we consider extremely unlikely.
 
\subsection{The [C~I]/CO intensity ratio}
\label{ratios}

As discussed in \S\ref{intro}, the strong interaction between UV
fields and the ISM in low metallicity environments leads us to expect
that the photodissociation products of CO will be enhanced in these
systems. Because low metallicity systems require a larger column of
molecular gas to achieve the UV extinction at which CO begins to form
($A_V\sim3$), we expect an increase in the [C~I]/CO intensity ratio in
metal-poor clouds compared with molecular clouds in the Milky Way
(Bolatto et al. 1999\markcite{BJI99}).

There are few observations of [C~I] in systems of low metallicity.
Wilson (1997)\markcite{WI97} measured the atomic carbon emission from
four individual molecular clouds in M33, a galaxy with a well studied
metallicity gradient. Wilson finds [C~I]/CO integrated intensity
ratios in the range 0.18---0.04 with an average $\simeq0.1\pm0.03$
(intensities in erg s$^{-1}$ cm$^{-2}$ sr$^{-1}$).  This ratio is very
similar to that of Galactic clouds, and there is no apparent
enhancement in the two clouds with lower metallicity. It is, however,
difficult to interpret this result because of the lack of other
diagnostics in these sources (e.g., [C~II] measurements).

Using the convolved interferometric CO ($J=1\rightarrow0$) data
(Wilson \& Reid 1991\markcite{WR91}) we have computed the I$_{\rm
[CI]}$/I$_{\rm CO}$ ratio for our 6 pointings (Table \ref{tabco}). The
average integrated [C~I]/CO intensity ratio for IC~10-SE is
$0.23\pm0.03$ when the intensities are in Rayleigh-Jeans observational
brightness units of K km s$^{-1}$, or equivalently $18\pm2$ when the
intensities are expressed in units of erg s$^{-1}$ cm$^{-2}$
sr$^{-1}$. Notice that this ratio is in principle an upper limit,
since part of the CO flux may be resolved out by the
interferometer. This result is similar to the [C~I]/CO$\approx20$
ratio observed by Stark et al. (1997)\markcite{ST97} towards N159-W in
the Large Magellanic Cloud, and $\sim50$\% larger than the COBE FIRAS
Galaxy-wide ratio of $\approx13$ (Wright et al. 1991\markcite{W91}).
Such similarity is surprising in light of the factor of 4 in
metallicity spanned by these sources, and the disparity in their
[C~II] intensities.  Bolatto et al. (1999\markcite{BJI99}) developed a
model for the [C~I]/CO intensity ratio as a function of
metallicity. According to this model, a roughly constant [C~I]/CO
ratio is difficult to understand if most of the neutral carbon is
predominantly produced by UV and is thus located in the PDR, the size
of which must obey relatively simple scaling laws with metallicity.
In the context of this model we can only understand a constant
[C~I]/CO ratio if most of the C$^0$ is not produced by
photodissociation but by other chemical processes (e.g., an enhanced
charge exchange reaction C$^+$ + S $\rightarrow$ C + S$^+$), either in
the surfaces of CO clumps or well mixed within the CO cores. Otherwise
we expect a sharp increase in the [C~I]/CO ratio with decreasing
metallicity, mostly owed to the decreasing size of the CO cores in the
clumps. Such an increase is apparent in the data for the [C~II]/CO
ratio, not a surprise since the dominant C$^+$ source is clearly UV
photodissociation. This dichotomy in the behaviour of C$^+$ and C$^0$
points to, in our opinion, different dominant origin mechanisms for
C$^+$ and C$^0$.

\begin{figure}
\plotfiddle{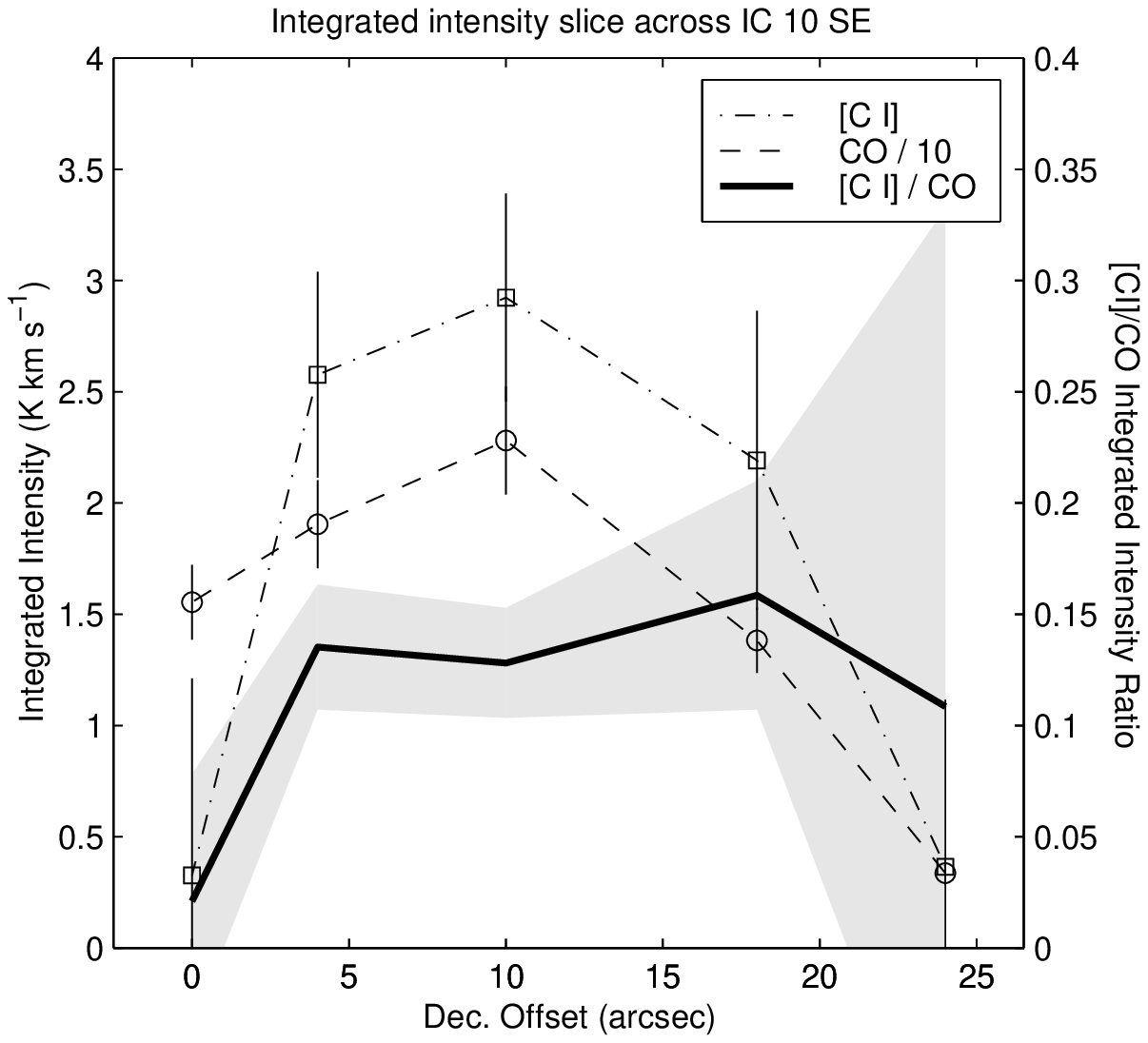}{2.5in}{0}{65}{65}{-205}{-140}
\plotfiddle{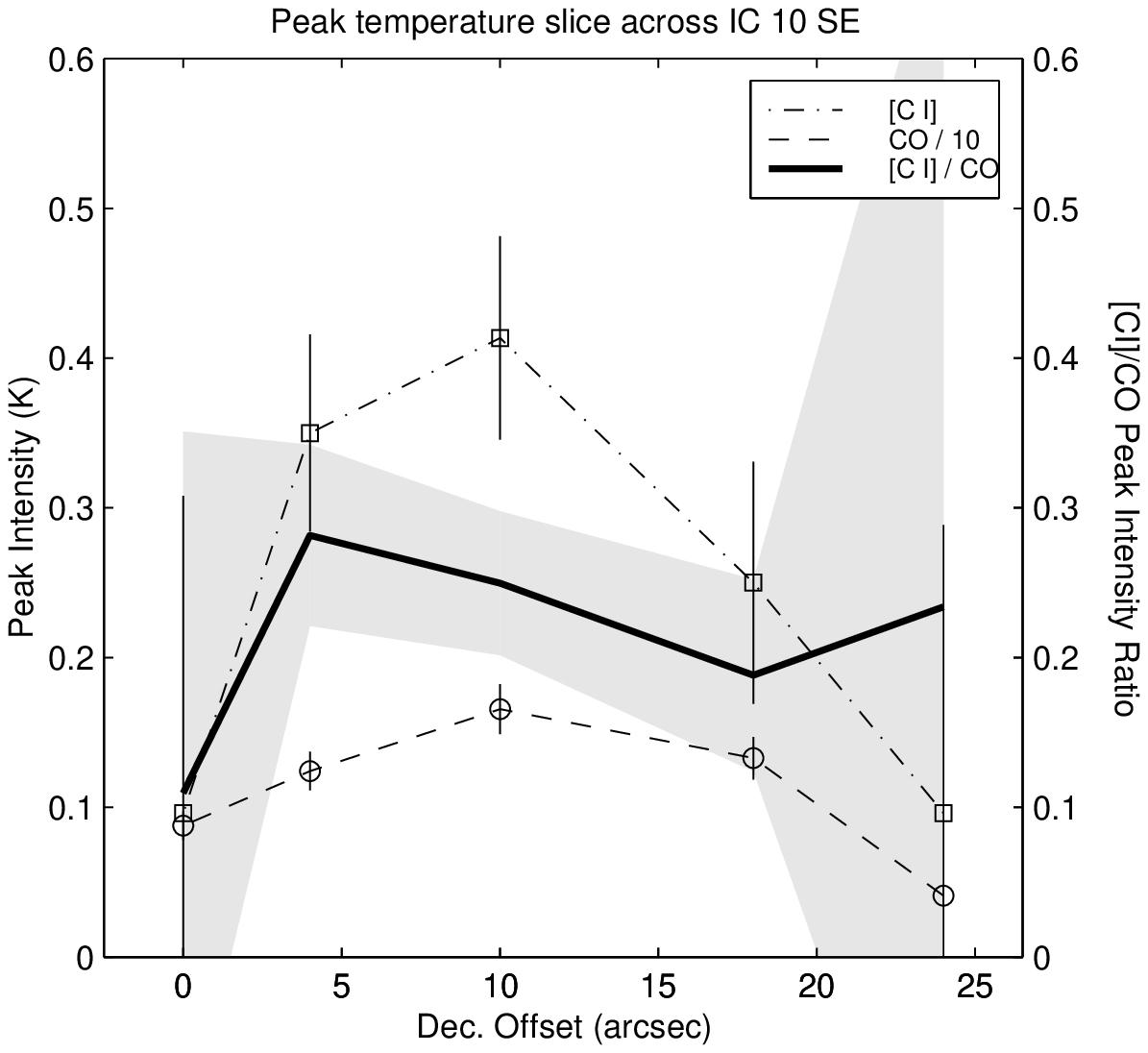}{2.5in}{0}{65}{65}{-205}{-165}
\figcaption[fig4a.eps,fig4b.eps]{[C~I] and CO ($J=3\rightarrow2$) data
in the slice across IC~10-SE.  {\em (Top)} The squares correspond to
the [C~I] integrated intensity (scale is on the left vertical axis)
and the circles correspond to the CO integrated intensity, scaled down
by a factor of 10. The ratio ${\rm I_{[CI]}/I_{CO}}$ along the slice
is shown by the solid line (scale on the right vertical axis). The
$1\sigma$ error bars have been computed including the internal errors
as well as a 10\% $1\sigma$ uncertainty in the calibration.  {\em
(Bottom)} Same as above but for the peak intensities along the slice.
Both plots are compatible with an essentially constant ratio across
IC~10-SE.\label{slice}}
\end{figure}

\begin{figure}
\plotone{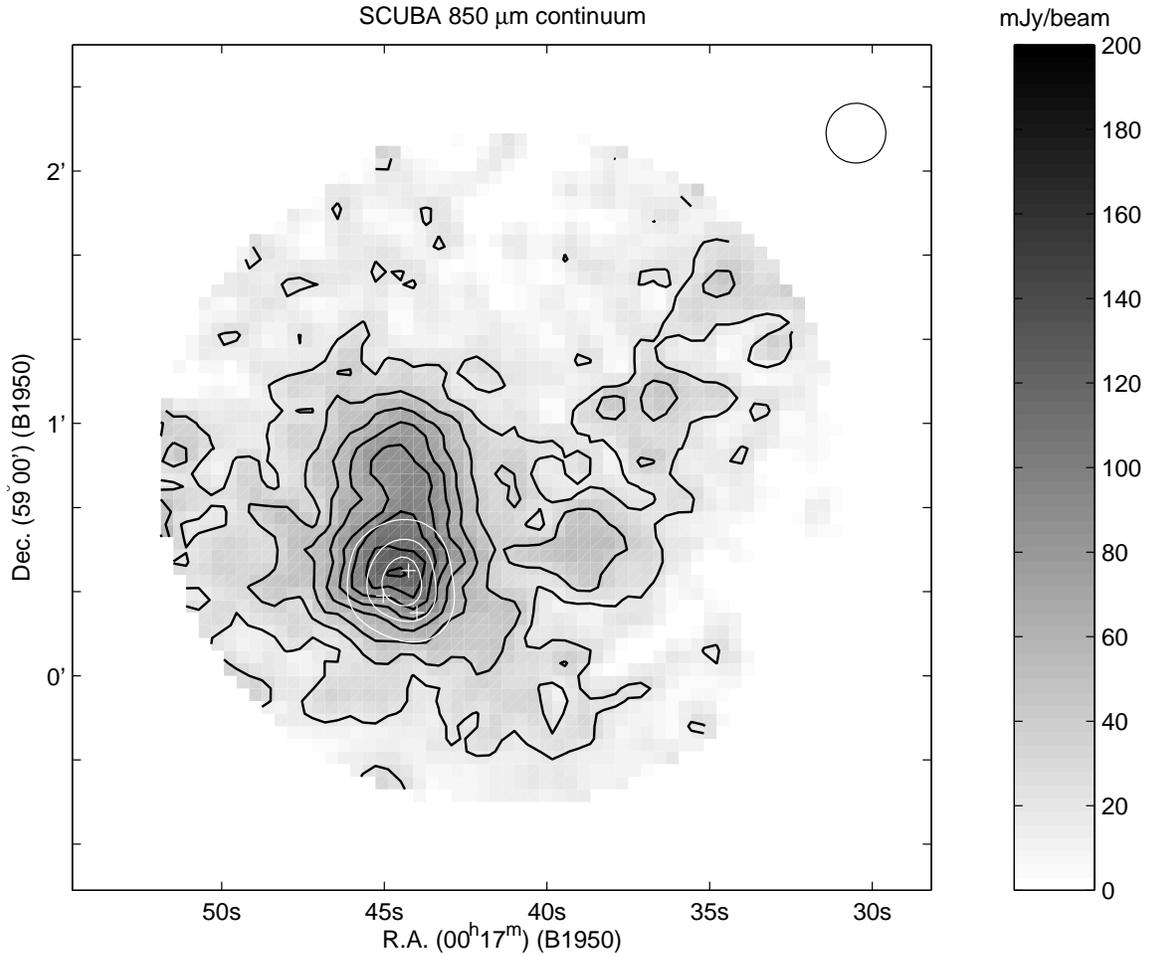} \figcaption[fig5.eps]{OVRO CO ($J=1\rightarrow0$)
contours (white) overlayed on our SCUBA 850 \micron\ continuum map
(grayscale). The circle in the upper right corner illustrates the size
of SCUBA's beam. The contour levels are 23 to 128 in steps of 15 mJy
per beam ($1\sigma=7.5$ mJy/beam).\label{scuba}}
\end{figure}

Figure \ref{slice} shows the variation in the [C~I] to CO
($J=3\rightarrow2$) ratio across the IC~10-SE complex. Both the ratio
of integrated intensities and the ratio of peak intensities remain
constant along the slice within the errors. This indicates that [C~I]
and CO emission are coextensive and we are not resolving the C$^0$/CO
transition, not a surprise given that the JCMT [C~I] beam subtends
$\approx44$ pc at the distance of IC~10.  Towards the ends of the
strip, at potentially interesting places to observe changes in the
[C~I]/CO ratio, the signals are very weak and we lack the
signal-to-noise necessary to measure this ratio.  In the case of
IC~10-SE the peak intensity ratio is on average about twice the
integrated intensity ratio, reflecting the difference in linewidths
discussed in \S\ref{comparison}.

Using the observed intensities of [C~II], [C~I] and CO we can compute
the relative abundances of the three dominant forms of carbon in this
source.  To this effect we will use Eqs. 20 from Bolatto et al.
(1999)\markcite{BJI99}, which are accurate to within factors of
$\sim2$ for [C~I] and [C~II] under a wide range of conditions. The
observed [C~I] and CO ($J=1\rightarrow0$) integrated intensities for
MC1 are ${\rm I_{[C~I]}} \cong 3.6\times10^{-7}$ and ${\rm I_{CO}}
\cong 2.2\times10^{-8}$ erg cm$^{-2}$ s$^{-1}$ sr$^{-1}$, measured in
a 10.8'' beam. The [C~II] intensity observed by Madden et
al. (1997)\markcite{MA97} is ${\rm I_{[C~II]}} \cong 7.5\times10^{-5}$
erg cm$^{-2}$ s$^{-1}$ sr$^{-1}$, albeit in a 55'' beam. The resulting
column densities are ${\rm N_{CO}}\cong1.3\times10^{18}$, ${\rm
N_{C^0}}\cong4.7\times10^{16}$, and ${\rm
N_{C^+}}\cong5.9\times10^{16}$ cm$^{-2}$. Petitpas \& Wilson
(1998)\markcite{PW98} obtained a somewhat larger column density ${\rm
N_{CO}}\cong6\times10^{18}$ cm$^{-2}$ from a multiline excitation
analysis, albeit in a larger beam (22'') that only partially contains
MC1 (it is displaced 14'' from our position).  Furthermore, using the
total column density of hydrogen derived from the dust continuum
measurements discussed in the following section (c.f., Table
\ref{tabsol}) the C relative abundance is ${\rm N_C/N_H}\simeq
1.3\times10^{18}/2.8\times10^{22} = 4.6\times10^{-5}$. This is a
factor of 4 smaller than the C abundance in a typical Milky Way cloud
like $\zeta$Oph (Duley \& Williams 1984)\markcite{DW94}, which is
consistent with the metallicity of IC~10-SE.

Therefore, unless there are strong local enhancements of ${\rm
N_{C^+}}$, CO is the dominant form of carbon in the molecular peaks of
IC~10-SE. It has been found, however, that in general CO is a poor
tracer of molecular gas in dwarf galaxies as it is discussed in
\S\ref{intro}. For example, Mochizuki et al.  (1994\markcite{MO94})
find that the distribution of [C~II] emission in the Large Magellanic
Cloud is very different from the CO.  Is this the case too in IC~10?
IC~10-SE is the molecular peak of this galaxy and one of the few
regions with enough extinction to form CO clouds (e.g., Becker
1990\markcite{BE90}).  The measurements by Madden et
al. (1997\markcite{MA97}) show other places in IC~10 with very large
I$_{\rm [CII]}$/I$_{\rm CO}$ ratios (e.g., position D is a [C~II] peak
with no known CO counterpart).  Thus, although locally CO is
preponderant in IC~10-SE, C$^+$ is likely to be the globally dominant
form of carbon in this galaxy.

\subsection{The dust continuum}
\label{dustcont}

Dust grains heated by starlight are the main source of the FIR
continuum emission. Since dust is also the source of UV extinction, it
determines the distribution of the photodissociated gas. The thermal
dust continuum contains information about the total mass of molecular
gas and the radiation field that is heating the grains, which can be
compared with estimates obtained by other methods. Finally, the
emissivity law of the dust particles yields information on the
composition of the dust grains and their size distribution. In this
section we present the first determination of the submillimeter dust
emissivity in a metal-poor system.

\begin{figure}
\plotone{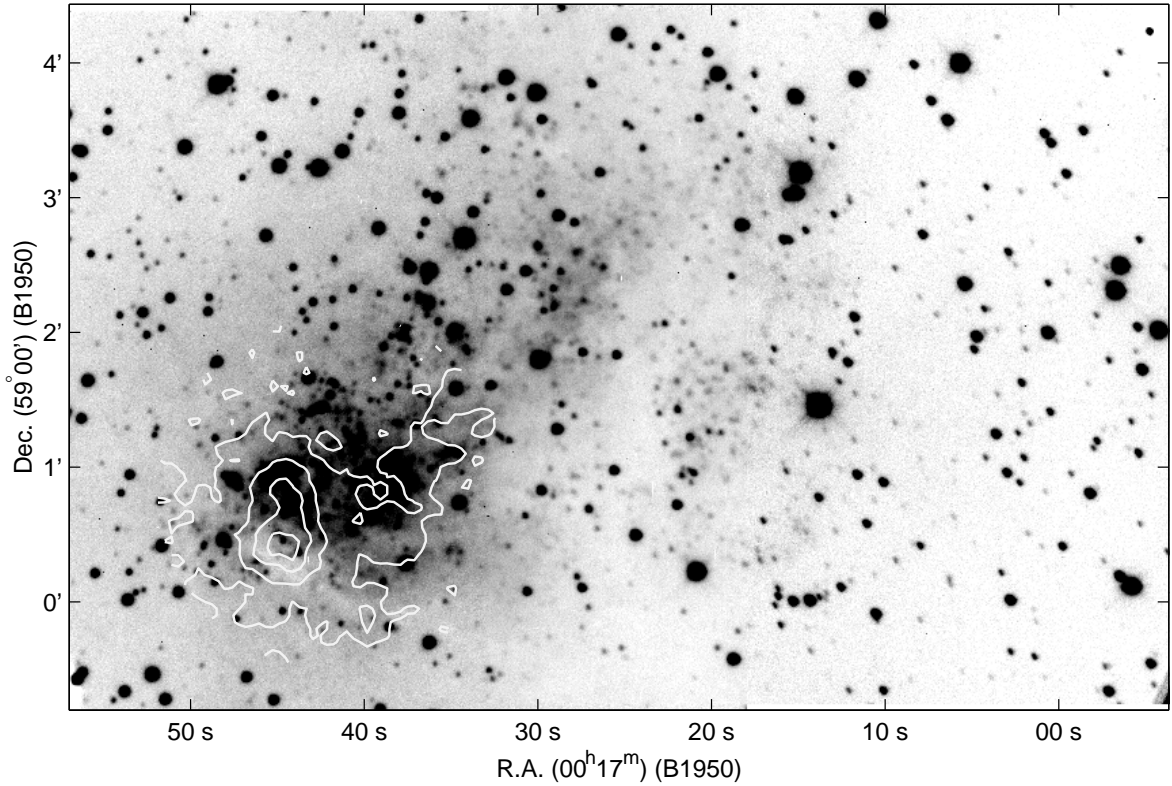} \figcaption[fig6.eps]{IC~10 in a B band mosaic
obtained at the Lowell Observatory 72'' Perkins telescope in Anderson
Mesa, Arizona, with the SCUBA map overlayed. The dust continuum is
strongest along an obscuration lane that corresponds to the peak of
the H~I map of Shostak \& Skillman (1989)\protect\markcite{SS89}.  The
void region near 00$^{\rm h}$17$^{\rm m}$25$^{\rm s}$,
59$^\circ$01'30'' corresponds to hole no. 2 in the same paper, while
the empty region close to 00$^{\rm h}$17$^{\rm m}$20$^{\rm s}$,
59$^\circ$02'30'' is probably associated with another of the H~I
peaks.\label{blue}}
\end{figure}

Figure \ref{scuba} shows the thermal dust continuum at 850 $\mu$m,
imaged with SCUBA down to a $1\sigma$ sensitivity of $\sim7.5$
mJy/beam.  The peak of the emission is coincident with the molecular
cloud complex IC~10-SE, with a northern extension and a fainter SE-NW
ridge of emission both associated with cluster of H~II regions and in
overall shape and dimensions very similar to the single dish CO
J=$1\rightarrow0$ map (Becker 1990\markcite{BE90}) and the 6 cm
radio-continuum observations of Yang \& Skillman (1993). Figure
\ref{blue} shows the dust continuum overlayed on a deep optical (B
band) picture of the galaxy.  Not surprisingly, the peak of SCUBA
emission is associated with an obscuration lane situated very close to
the center of star formation activity in IC~10.

\begin{figure}
\plotone{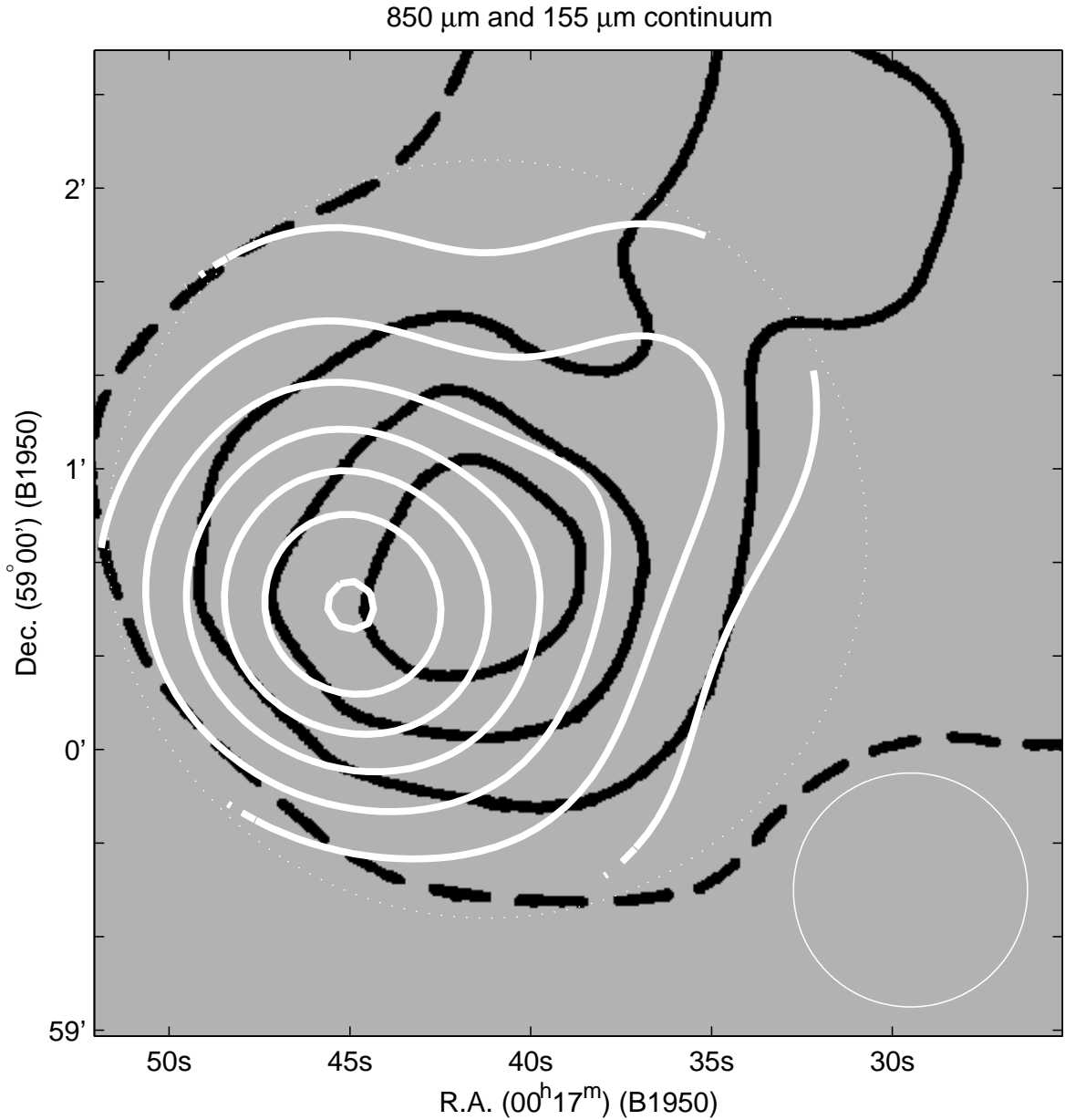} \figcaption[fig7.eps]{Previous KAO 155 \micron\
observations (Thronson et al. 1990\protect\markcite{TH90}, black
contours) compared with our SCUBA 850 \micron\ map convolved to a 50''
beam size (white contours). The displacement is most likely due to
pointing discrepancies with the KAO. The convolved SCUBA contour
levels are 100 to 700 in steps of 100 mJy per 50'' beam.  The contours
in the 155 \micron\ data are 5, 10 and 15 Jy per 50'' beam.  The
dotted circle shows the approximate area mapped by SCUBA, while the
white thin circle in the lower right corner shows one 50''
beam.\label{kao}}
\end{figure}

After convolving these observations to the resolution of previous FIR
maps of this source ($\sim50''$ HPBW, equivalent to $\sim200$ pc at
the adopted distance; Thronson et al. 1990\markcite{TH90}), we find
that the general morphology is well preserved although there is a
displacement of order 30'' between the 95 or 155 $\mu$m KAO data and
the 850 $\mu$m SCUBA map (Fig. \ref{kao}).  Although the estimated
$1\sigma$ positional uncertainty of the KAO data is 10'', we consider
that the observed displacement is an artifact of pointing.  Notice
that while the CO, C$^+$ and 850 $\mu$m peaks are all coincident, the
best correlation between the FIR map in Thoronson et al. and the C$^+$
and CO data is obtained only after such an offset is applied to the
FIR map (Madden et al. 1997\markcite{MA97}).  Unfortunatly the IRAS
HIRES maps of this region cannot be used to further clarify this
point, since their positional accuracy is only about 30''
(incidentally, the HIRES peak falls between the KAO and the SCUBA
positions).  The experience at KAO suggests that, while uncommon, such
pointing errors are entirely possible. This is especially true during
the observation of faint irregular galaxies when it is impossible to
autoguide on the source itself (Harper 1999), as was the case for
IC~10. Therefore, we consider that the 30'' displacement is most
likely a registration offset caused by an error in the KAO pointing,
and in our continuum analysis we use the peak values of the source in
each map.  Using the nominal FIR values at the position of the 850
$\mu$m peak ($S_{95}=17$ Jy, $S_{155}=14$ Jy) only causes a moderate
shift in the temperature of the solution, without altering the
conclusions (the solution for the free-free corrected 95, 155, 850,
and 1300 $\mu$m points would then be $T\simeq38$ K, $\tau_{\rm
mm}\simeq6.8\times10^{-5}$, $\beta\simeq0.6$. The IRAS HIRES point at
60 $\mu$m cannot be used because of the uncertainty in its position).

To compare our submillimeter measurements with the FIR data it is
necessary to convolve them to a common angular resolution. This is a
problem for the 1350 $\mu$m data, for which we have only a single
pointing with HPBW 18''. To overcome this problem we include data from
a 1.3 mm continuum map of IC~10 taken at IRAM. The corresponding flux
density of this region is 280 mJy in a 50'' beam (Wild 1998). The
FIR/submillimeter measurements are summarized in Table \ref{tabcont}.

For a modified blackbody (a graybody) the opacity $\tau_\lambda$ is a
function of wavelength of the form

\begin{equation}
\tau_\lambda = \left({\lambda_0\over\lambda}\right)^\beta
\end{equation}

\noindent where $\beta$ is the graybody emissivity exponent, and
$\lambda_0$ is the wavelength at which the emission becomes optically
thick. Accordingly, the thermal emission from a graybody can be
expressed as

\begin{equation}
S_\nu = \Omega B_\nu(T)
\left(1-e^{-({\lambda_0\over\lambda})^\beta}\right)
\end{equation}

\noindent where $S_\nu$ is the observed flux density, $\Omega$ is the
source solid angle, and $B_\nu(T)$ is Planck's function.  In the
optically thin regime a simultaneous determination of the source size
and opacity is impossible, and we can only determine the product
$\Omega\tau_\lambda$. We set $\Omega=\Omega_{beam}=6.7\times10^{-8}$
sr$^{-1}$ and the opacity thus determined is an average over the beam.

The dust opacity in the FIR-submillimeter regime is generally very
small ($\tau\ll 1$). In the optically thin limit the opacity is
related to the hydrogen column density by $\tau_\lambda = b\,
\sigma_\lambda N_{\rm H}$ and thus

\begin{equation}
\frac{S_\nu}{\Omega} = b\, \sigma_\lambda N_{\rm H} B_\nu(T)
\end{equation}

\noindent where $\sigma_\lambda$ is the dust opacity per hydrogen atom
per cm$^{2}$ in the diffuse ISM (the canonical dust cross-section),
$b$ is a numerical factor (see below), and the hydrogen column density
is $N_{\rm H}=N{\rm(H)}+2N{\rm(H_2)}$. The relationship between the
opacity and the hydrogen column density can be expressed as

\begin{equation}
\tau_\lambda \simeq b\, \frac{Z}{Z_\odot}\, \sigma_{\rm mm}
\lambda_{\rm mm}^{-\beta} N_{\rm H}
\end{equation}

\noindent where $Z/Z_\odot$ is the metallicity of the gas relative to
the Sun, $\sigma_{\rm mm}$ is the value of $\sigma_\lambda$ at a
wavelength of 1 mm and solar metallicity, $\lambda_{\rm mm}$ is the
wavelength in mm and it is implicitly assumed that the dust-to-gas
ratio is proportional to the metallicity. The numerical factor $b$ is
of order unity and depends on the grain environment: $b\approx1$ for
the dust in the diffuse interstellar medium considered by Draine \&
Lee (1984)\markcite{DL84}, $b>1$ for dust in deeply embedded sources
where ice mantles form on the grain surfaces, $b<1$ in environments
where grain destruction is occurring (Mezger, Wink, \& Zylka
1990\markcite{ME90}; Braine et al. 1997\markcite{BR97}). Ossenkopf \&
Henning (1994)\markcite{OH94} have modeled the opacity of interstellar
grain aggregates in a variety of physical conditions applicable to
protostellar cores (i.e., highly embedded sources). They find that for
such sources the submillimeter opacity per unit dust mass can be a
factor of 5 larger on average than in the diffuse interstellar medium,
implying $b\sim5$.  The increased opacity in these environments is
caused by the formation of thick ice mantles and the coagulation of
fluffy particle aggregates (e.g., Fogel \& Leung,
1998\markcite{FL98}). The theoretical estimate for the dust opacity
per hydrogen atom is $\sigma_{\rm mm}=7\times10^{-27}$ cm$^2$ in
diffuse gas (Draine \& Lee 1984\markcite{DL84}), while observations
place this number a factor of 1.2---1.7 higher in a wide range of
objects (Braine et al. 1997\markcite{BR97}, and references therein).

\begin{figure}
\plotone{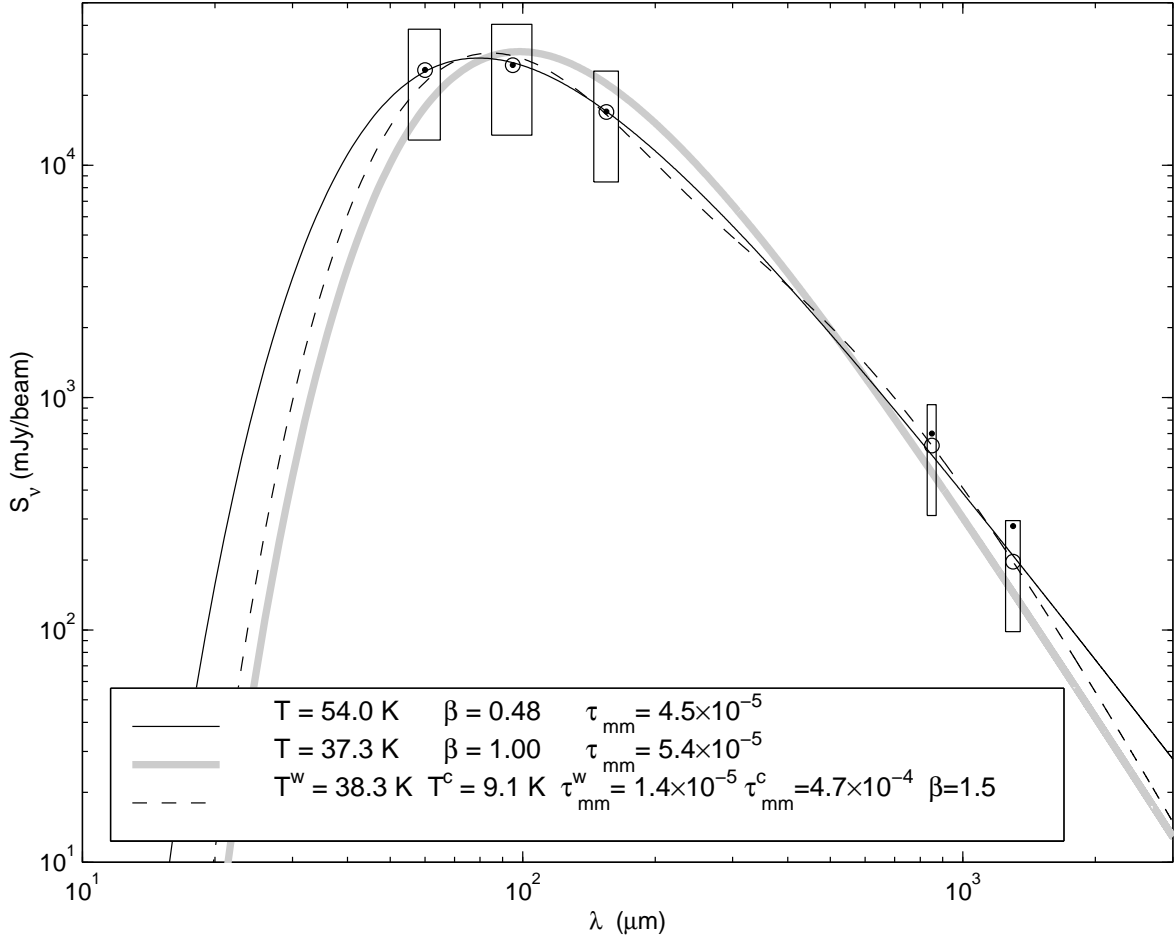} \figcaption[fig8.eps]{Flux density distribution of
IC~10-SE in the FIR-submillimeter region, showing the fits to the 60,
95, 155, 850 and 1300 $\mu$m free-free corrected data (small
circles). The original measurements are shown here by black dots.  The
$3\sigma$ error boxes are those in Table \protect\ref{tabcont}.  The
main graybody solution (black solid line), with emissivity exponent
$\beta\sim0.5$, is shown here together with the best $\beta=1$ fit and
the two-temperature solution for $\beta=1.5$.
\label{bbsol}}
\end{figure}

Figure \ref{bbsol} shows the measured flux density with the
corresponding modified blackbody solutions.  The solutions were
computed using the 60, 95, 155, 850 and 1300 $\mu$m points listed in
Table \ref{tabcont}.  The best-fit single temperature graybody
solution to the flux density has a very shallow slope in the
Rayleigh-Jeans limit, corresponding to a FIR-submillimeter emissivity
exponent $\beta\sim0.3$ at a temperature $T\simeq60$ K with an opacity
at 1 mm $\tau_{\rm mm}\simeq5\times10^{-5}$ (notice that this solution
is slightly different from the median solution in the Monte Carlo
listed in Table \ref{tabsol}).  Fitting the data with a blackbody
produces $T\sim85$ K and $\tau_{\rm mm}\sim4 \times10^{-5}$. In fact,
the slope derived from the 850 and 1300 $\mu$m measurements alone is
$\lambda^{-(2+\beta)}\simeq\lambda^{-2.16}$, very close to a
blackbody. This extremely shallow emissivity is unusual. Grain
emissivity exponents for different materials are thought to be in the
range $1<\beta<2$, corresponding to amorphous ($\beta\simeq1$) and
metallic or crystalline ($\beta\simeq2$) structure (e.g., Draine \&
Lee 1984\markcite{DL84}; Tielens \& Allamandola 1987\markcite{TA87};
Mennella, Colangelli, \& Bussoletti 1995\markcite{ME95}).
Astronomical measurements generally confirm this result for both
Galactic (e.g., Knapp, Sandell, \& Robson 1993\markcite{KN93}) and
extragalactic sources (e.g., Hughes, Gear, \& Robson
1994\markcite{HU94}).

The grain emissivity may be influenced by environmental
characteristics, mainly the interstellar radiation field.  IC~10-SE is
in close proximity to the most luminous H~II region in IC~10 (region
111, Hodge \& Lee 1990\markcite{HL90}).  This region has an H$_\alpha$
flux of $2.8\times10^{-13}$ erg s$^{-1}$ cm$^{2}$, and consequently a
luminosity $L_{H_\alpha}\sim4.5\times10^{37}$ erg s$^{-1}$ (assuming
an extinction correction $A_{H_\alpha}=0.8$ and $D=820$ kpc; Thronson
et al. 1990\markcite{TH90}; Wilson et al. 1996\markcite{WI96}), or
about four times as luminous as the Orion nebula (Kennicutt
1984\markcite{KE84}).  We can estimate the average radiation field in
this region by assuming that all the starlight is reradiated in the
FIR.  Using our graybody solution, the observed FIR surface brightness
is $S_{\rm FIR}\approx1.5\times10^{-9}$ erg s$^{-1}$ cm$^{-2}$
beam$^{-1}$, resulting in a flux $F_{\rm FIR}\approx0.30$ erg s$^{-1}$
cm$^{-2}$. This is equivalent to an average radiation field
$\chi_{uv}\simeq190\chi_0$ over a region 200 pc in diameter, where
$\chi_0$ is the standard radiation field in the vicinity of the Sun
($\chi_0 = 1.6\times10^{-3}$ erg s$^{-1}$ cm$^{-2}$; Habing
1968\markcite{HA68}). Most of the submillimeter emission comes from a
region $\sim22$'' ($\sim90$ pc) wide. Assuming that all the flux
originates in that region, its radiation field is
$\chi_{uv}\sim900\chi_0$. This value is probably an underestimate,
since we are neglecting: {\em 1)} the contribution of the
stochastically heated small grains, which will produce excess
radiation shortwards of 100 $\mu$m and increase the integrated FIR
flux, and {\em 2)} the possibility that any form of radiation other
than FIR is escaping the region (i.e., we assume 100\% UV to FIR
conversion efficiency).  Nevertheless, this estimate is about a factor
of four higher than the average radiation field in the Orion region
where $\chi_{uv}\sim250$ (over scales of tens of parsecs; Stacey et
al. 1993\markcite{ST93}) and agrees very well with estimates for
$\chi_{uv}$ based on [C~II] data for the same region (Madden et
al. 1997).

Contamination by free-free radiation is a possible cause for a shallow
long wavelength slope in the graybody solution. Optically thin thermal
free-free emission has a flat spectrum ($S_\nu\propto\nu^{-0.1}$) that
we can remove from the longer wavelength data to increase the slope of
the graybody in the Rayleigh-Jeans limit.  To obtain $\beta$=1---2
using solely the two longest frequency measurements, the contribution
from the free-free radiation should be $S_\nu\simeq$120---190 mJy at
$\lambda\sim1$ mm.  One way to assess how much free-free is
contaminating our fluxes is to simultaneously fit the graybody and the
free-free emission (4 free parameters, 5 data points).  Doing so
results in only 7 mJy free-free at 1 mm, therefore we expect a rather
small free-free contamination in our continuum data. Klein \& Gr\"ave
(1986)\markcite{KG86} found that the galaxy-wide flux density for
IC~10 at 6.3 cm is 222 mJy with a spectral index $\alpha\simeq-0.33$,
slightly steeper than purely thermal bremsstrahlung radiation. The
spectral index flattens to $\alpha=-0.1$--- $-0.2$ at the location of
the two emission peaks, one of which appears to be associated to
IC~10-SE (peak ``B'' for Klein \& Gr\"ave).  Assuming $\alpha=-0.15$,
these numbers predict a free-free contribution to the spectral density
$S_\nu\simeq120$ mJy at 1 mm, with roughly \twothirds\ of the flux
originating in the southernmost component identified with IC~10-SE.
Therefore we assume $S^{ff}_{\nu}\approx80\,\lambda_{\rm mm}^{0.15}$
mJy.  Removing this contribution from the observed fluxes will raise
$\beta$ to about 0.5 when fitting all data points.  Thus, we find that
although contamination by bremsstrahlung radiation should be taken
into account when deriving the graybody solution, it is not enough to
bring the graybody emissivity exponent up to $\beta=1$.

\begin{figure}
\plotone{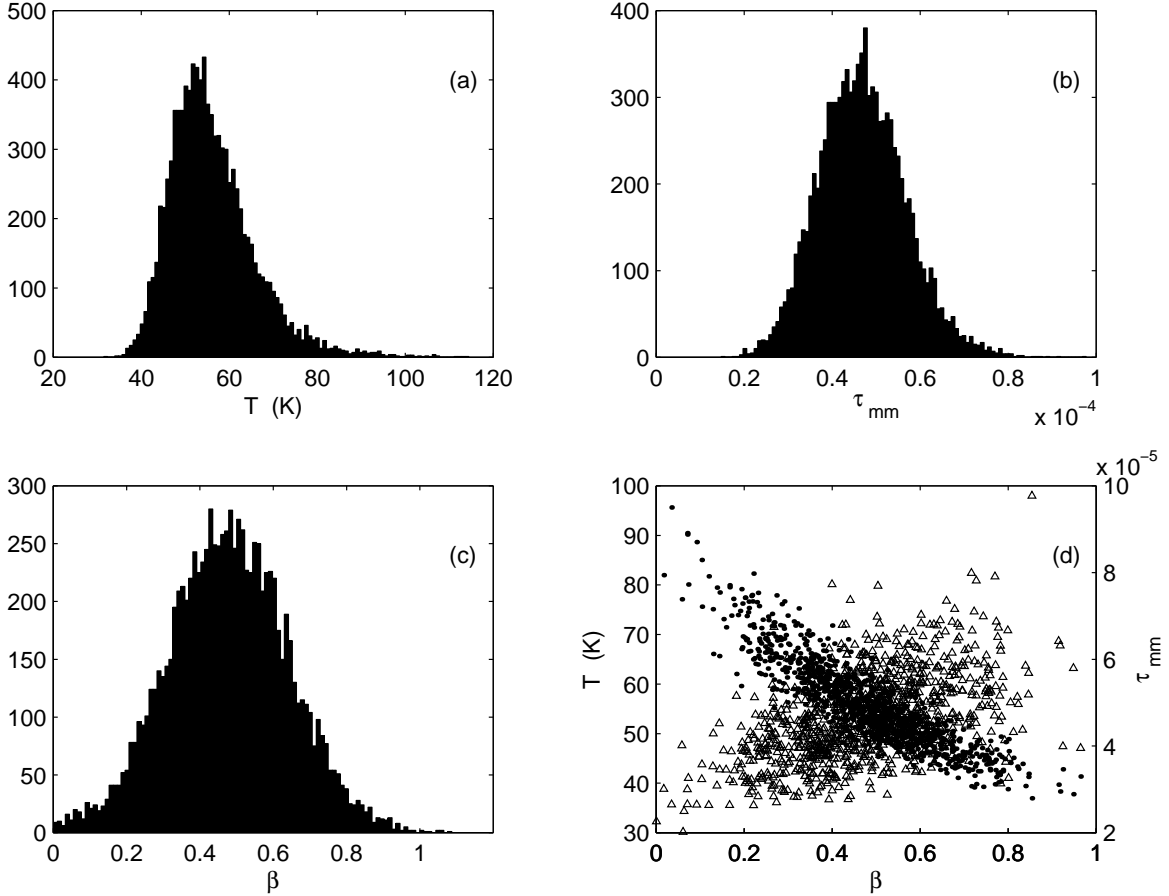} \figcaption[fig9.eps]{$10^4$ point Monte Carlo
error analysis for the graybody solution derived using the combined
KAO, SCUBA and IRAM measurements described in the text.  All the
solutions with $\beta<0$ (about 19\%) were considered unphysical and
removed from the plots and the distributions before computing the
median and the error estimates.  The histograms (a), (b), and (c) show
the distribution of the solutions for the graybody parameters: $T$,
$\tau_{\rm mm}$ and $\beta$. The scatter plot (d) shows about 10\% of
the solutions for $T$ (dots) and $\tau_{\rm mm}$ (triangles) plotted
against their emissivity exponent, $\beta$.  Thus, high temperature
and low $\tau_{\rm mm}$ occur for low $\beta$.  The mm opacity
$\tau_{\rm mm}$ is relatively insensitive to the values of the other
parameters in the graybody solution.\label{bbloc}}
\end{figure}

How much can we trust our determination of the submillimeter
emissivity?  We have performed a $10^4$ points Monte Carlo analysis on
the free-free corrected measurements, assuming independent gaussian
errors with $3\sigma$ deviations according to Table \ref{tabcont} (the
assumed $3\sigma$ deviations in flux correspond to 50\% calibration
errors). The results of this analysis are shown in Figure \ref{bbloc}
and allow us to discard emissivity exponents $\beta\gtrsim1$ as
statistically very improbable.  Another possible source of error is a
bad data point, or a systematic calibration difference among the
different telescopes.  Eliminating the data from any one of the
telescopes listed in Table \ref{tabcont}, however, still results in an
emissivity exponent very close to zero for the fit to the remaining
data. Both KAO points, for example, need to be raised by a factor of
$\sim2$ to produce $\beta\sim1$ when fitting all five
measurements. Although this is not impossible, we find unlikely that a
bad data point or a systematic calibration difference is causing our
low $\beta$ result.

A third possibility is the presence of a substantial mass of very cold
dust in the line of sight. Since we have five measurements, we can fit
a two temperature component graybody with an arbitrary emissivity. The
results are summarized in Table \ref{tabsol}.  To determine if an
important component of cold dust is a possibility we will compute the
mass of cold dust and compare it to virial estimates for the mass of
this region.  Although a large mass of cold dust is consistent with
the shape of the submillimeter continuum, such a cold dust component
would contain so much mass that it is inconsistent with the observed
column density of hydrogen.  The H~I distribution for IC~10 peaks in
the IC~10-SE region, at a column density of $4.1\times10^{21}$
cm$^{-2}$ imaged with a 30'' beam (Shostak \& Skillman 1989). Single
dish CO observations (Becker 1990\markcite{BE90}), when corrected by
metallicity effects (Wilson 1995\markcite{WI95}), imply a molecular
hydrogen column density $N$(H$_2$)$\simeq9.3\times10^{21}$ cm$^{-2}$
for IC~10-SE (see also Petitpas \& Wilson, 1998\markcite{PW98}).  The
total column density of hydrogen is thus $N_{\rm H}=N_{\rm HI}+2
N_{\rm H_2}\simeq2.3\times10^{22}$ cm$^{-2}$.  This figure agrees very
well with the estimate based on the low emissivity graybody models
setting $b\sim1$ (c.f., Table \ref{tabsol}), and is smaller than the
column densities predicted by the two temperature models by about {\em
a factor of 10}. As we discussed before, highly embedded sources may
have $b$ as large as 5, which will diminish this discrepancy to a
factor of about 2.  The gas intermixed with such a cold dust component
should be molecular and emit strongly in CO $J=1\rightarrow0$ but
little or nothing in the $J=2\rightarrow1$ transition. Its emissivity
will probably be about a factor of 4 smaller than the molecular gas
associated with the warm component (i.e., the ratio of dust
temperatures in the two components), and thus could be conceivably
``masked'' by the warmer gas associated with active star-forming
regions. If we adopt the $\beta=1.5$ two-temperature solution in Table
\ref{tabsol} and proceed to correct down the expected $N_{\rm H}$
column density by a factor of $5\times4=20$, we obtain a cold CO
($J=1\rightarrow0$) emissivity a factor of $\sim2$ larger than that of
the warm CO. Therefore, we should see about 3 times more flux in CO
($J=1\rightarrow0$) than in CO ($J=2\rightarrow1$).

We will compare now this expectation with observations.  A multiline
excitation analysis of IC~10-SE has been performed by Petitpas \&
Wilson (1998)\markcite{PW98}.  No unique solution was found by the
authors, which they attribute to the presence of two density
components. Another possible explanation, however, would be the
existence of two temperature components.  The observed main beam line
temperature ratio is $^{12}$CO
($J=2\rightarrow1$)/($J=1\rightarrow0$)=$0.64\pm0.05$, somewhat
smaller than the typical line ratio for $^{12}$CO in Milky Way clouds
(about 0.8), but not as low as our reasoning in the previous paragraph
leads us to expect.  This is, however, further complicated by the fact
that $^{12}$CO is most probably optically thick. A more useful probe
(but much more difficult to observe) would be an optically thin
isotopomer like C$^{18}$O.  There are also no signs of self absorption
in the $^{12}$CO ($J=1\rightarrow0$) spectra (e.g., Becker
1990\markcite{BE90}), as expected if some of the cold CO is located in
front of the warm CO along the line of sight.  Notice that this
hypothetical very cold gas component cannot be the source of the
excess [C~II] emission observed by Madden et al.
(1997)\markcite{MA97} since the 158 $\mu$m transition requires 91 K to
be excited. Hence, although we cannot conclusively discard the
possibility a substantial mass of cold dust with the present data, we
consider a two temperature model with a dominant cold dust fraction
very unlikely and conclude that the submillimeter emissivity exponent
of the dust in this environment is intrinsically small.

Thus, we have explored and find unlikely the possibility of
calibration errors, free-free radiation and cold dust as the plausible
causes of the low emissivity exponent that we observe in this source.
Such a shallow emissivity must then be intrinsic, and related to some
unique characteristic of IC~10-SE such as its low metallicity and high
radiation field. Calculations show that for sufficiently small
graphite grains the photon absorption cross-section $Q_{abs}/a$ is
independent of the grain size $a$ and temperature $T$, as electric
dipole absorption is the dominant mechanism (Draine \& Lee
1984\markcite{DL84}). For larger grains ($a\gtrsim0.5$ $\mu$m)
magnetic dipole absorption becomes important and $Q_{abs}/a$ shows a
strong temperature dependence for wavelengths longwards of 100 $\mu$m,
as well as a weaker dependence on $\lambda$ which translates into a
shallower emissivity (c.f., Figure 4b of Draine \& Lee
1984\markcite{DL84}).  Quantitatively, however, inspection of Figure
4b reveals that $\beta\sim1.5$ for the largest and hottest graphite
grains and $\beta\sim1.8$---2 otherwise (silicate grains and smaller
graphite grains), values typical for crystalline materials. There are
no similar calculations for amorphous carbonaceous materials, which
laboratory measurements show to have $\beta\sim0.8$---1.3 (Menella et
al. 1995\markcite{ME95}). Hence, although theory suggests that a small
$\beta$ is possible if small grains are absent and the dust size
distribution is dominated by large grains, there is no actual
quantitative support for $\beta\sim0.5$.  Observational evidence for
small grain destruction in starburst environments similar to IC~10-SE
has been found by several studies, for example: PAH destruction in the
core of M82 (Normand et al. 1995\markcite{NO95}), gas phase silicon
enhancement associated with silicate grain destruction by
supernovae-driven shocks (Lord et al.  1996\markcite{LO96}), FIR
color-UV field correlation caused by small grain evaporation (Telesco,
Decher, \& Joy 1989\markcite{TE89}; Boulanger et
al. 1988\markcite{BO88}), and metallicity effects on the FIR colors
attributed to small grain destruction (Sauvage, Thuan, \& Vigroux
1990\markcite{SA90}). In view of this, we consider possible that the
small emissivity exponent of the thermal dust continuum results from
small grain destruction occurring in an actively star-forming low
metallicity environment.

The continuum from the nucleus of M~82 has been studied in detail by
Hughes et al. (1994\markcite{HU94}), who found for it a normal
emissivity exponent ($\beta=1.3$) despite the high radiation field
($\chi_{uv}\sim4700$ in the central region, $\chi_{uv}\sim440$ over a
larger extent; Stacey et al. 1991\markcite{ST91}).  It is important to
keep in mind, however, that the metallicity of the nucleus of M~82 is
about 14 times larger than the metallicity of IC~10 (Alloin et
al. 1979\markcite{AL79}).  The ratio of $\chi_{uv}/Z$ is then $\sim3$
times larger for IC~10-SE versus the central region of M~82. Naively,
this would indicate a 3 times bigger ``effective'' radiation field
(i.e., corrected by dust extinction) in IC~10.  Without detailed
modelling of the physical processes involved in grain formation and
destruction, however, it is difficult to quantify the separate effects
of $Z$ and $\chi_{uv}$ on the grain size distribution.  Nevertheless,
it is important to point out that a high radiation field alone, such
as the one found in the cores of starburst galaxies, may not be
sufficient to produce as shallow an emissivity law as we observe in
IC~10.  Such a low $\beta$ may require both low metallicity and a high
radiation field.

This result, if confirmed, has potentially important implications for
the recently opened field of continuum observations of young galaxies
at high redshifts (e.g., Smail, Ivison, \& Blain 1997\markcite{SM97};
Barger et al. 1998\markcite{BA98}; Barger, Cowie, \& Sanders
1999\markcite{BA99}) since it would affect the shape of their spectral
energy distribution (SED) at long wavelengths. With equal $T$ and
$\tau_{\rm mm}$, a galaxy with smaller $\beta$ will be brighter in the
Rayleigh-Jeans portion of its continuum. Thus galaxy counting surveys
would tend to be dominated by such sources, and backwards
extrapolation from the current galaxy population using a standard SED
with $\beta=1.5$ may considerably underpredict the observed galaxy
counts at high redshifts.

\section{Summary}

We have observed [C~I], $^{12}$CO and $^{13}$CO (J=$3\rightarrow2$),
and dust continuum in the SE region of IC~10.  This is one of the few
submillimeter neutral carbon studies of a metal-poor system (Stark et
al. 1997\markcite{ST97}; Wilson 1997\markcite{WI97}).

We detected the [C~I] 609 \micron\ line towards four out of six
positions in IC~10-SE, at the $3\sigma$ level or higher.  All four
points share a similar value of I$_{\rm [CI]}/$I$_{\rm CO}\simeq
18\pm2$ (in erg s$^{-1}$ cm$^{-2}$ sr$^{-1}$). This ratio is similar
to that observed in the Orion region and most Milky Way sources
(I$_{\rm [CI]}/$I$_{\rm CO}\sim$ 10---15), and the N~159 region in the
Large Magellanic Cloud (I$_{\rm [CI]}/$I$_{\rm CO}\simeq20$), in spite
of these regions spanning a factor of 4 in metallicity and a factor of
3 in I$_{\rm [CII]}/$I$_{\rm CO}$.  The similarity of the I$_{\rm
[CI]}/$I$_{\rm CO}$ ratio among sources of disparate metallicity is
associated to the nature of the C$^+$/C$^0$/CO transition in molecular
clouds and possibly the structure of the clouds themselves (Bolatto et
al.  1999\markcite{BJI99}; Pak et al. 1998\markcite{PA98}).  We
believe that the observed trend of constant I$_{\rm [CI]}/$I$_{\rm
CO}$ ratio with decreasing metallicity and increasing I$_{\rm
[CII]}/$I$_{\rm CO}$ points to a non-PDR origin for most of the C$^0$
in molecular clouds. While the final settling of this question must
necessarily await for better statistics, the hints offered by these
observations are tantalizing.

The spectral energy distribution of the submillimeter continuum is
relatively shallow, resulting in a graybody emissivity exponent
$\beta\sim0.5$. It is unlikely that this small emissivity exponent is
caused by errors in the measurements, a miscalibration of one of the
telescopes, contamination of the thermal greybody spectrum by
free-free emission or a substantial amount of cold dust along the line
of sight, therefore we believe it is intrinsic.  Furthermore, the
standard hydrogen column density estimate based upon the dust
continuum emission agrees very well with the column density derived
using other methods.

The physical cause of the shallow emissivity may be the destruction of
small grains by UV radiation, which effectively penetrates farther
into the ISM in low metallicity environments with low dust-to-gas
ratios. This study of a nearby, low metallicity interstellar medium
actively forming stars has direct relevance to present and future
efforts to observe the young galaxy population at high redshifts.

\acknowledgements The authors wish to thank D. P. Clemens for useful
comments on the draft of this paper, as well as an anonymous referee
for a very thorough job that definitely improved this work.  The
research of A.D.B. and J.M.J. was supported in part by the National
Science Foundation through grant AST-9803065.  The research of
C.D.W. is supported through a grant from the Natural Sciences and
Engineering Research Council of Canada.  This research has made use of
NASA's Astrophysics Data System Bibliographic Services.  The JCMT is
operated by the Royal Observatory Edinburgh on behalf of the Particle
Physics and Astronomy Research Council of the United Kingdom, the
Netherlands Organization for Scientific Research, and the National
Research Council of Canada.

\newpage

\newpage

\newpage

\newpage

\newpage

\newpage

\newpage

\newpage

\newpage

\newpage

\begin{deluxetable}{lccr@{$\pm$}lr@{$\pm$}lr@{$\pm$}lr@{$\pm$}lcc}
\tablewidth{0pc} \scriptsize \tablecaption{Measured [C I] line
parameters for IC~10-SE} \tablehead{ & R.A.  & Dec &
\multicolumn{2}{c}{T$_{mb}$\ \tablenotemark{a,b}} &
\multicolumn{2}{c}{V$_{\rm LSR}$\ \tablenotemark{b}} &
\multicolumn{2}{c}{FWHM\ \tablenotemark{b}} &
\multicolumn{2}{c}{I$_{\rm [C I]}$\ \tablenotemark{a,c}} & RMS\
\tablenotemark{d} & $\tau_{int}$\ \tablenotemark{e} \\ & (B 1950) & (B
1950) & \multicolumn{2}{c}{(K)} & \multicolumn{2}{c}{(km s$^{-1}$)} &
\multicolumn{2}{c}{(km s$^{-1}$)} & \multicolumn{2}{c}{(K km
s$^{-1}$)} & (K) & (s)\\ } \startdata MC 1 & $00^{\rm h}17^{\rm
m}44\fs3$ & 59\arcdeg00\arcmin25\arcsec & 0.41 & 0.06 & -329.6 & 0.4 &
\ 6.9 & 1.0 & 2.92 & 0.37 & 0.13 & 5400\\ MC 2 & $00^{\rm h}17^{\rm
m}45\fs0$ & 59\arcdeg00\arcmin19\arcsec & 0.35 & 0.06 & -328.9 & 0.5 &
6.9 & 1.3 & 2.56 & 0.39 & 0.13 & 5400\\ MC 3 & $00^{\rm h}17^{\rm
m}44\fs0$ & 59\arcdeg00\arcmin15\arcsec & 0.21 & 0.06 & -330.6 & 1.1 &
9.6 & 2.7 & 2.23 & 0.50 & 0.13 & 4500\\ SL 1 & $00^{\rm h}17^{\rm
m}43\fs7$ & 59\arcdeg00\arcmin39\arcsec & \multicolumn{2}{c}{$\cdots$}
& \multicolumn{2}{c}{$\cdots$} & \multicolumn{2}{c}{$\cdots$} & 0.37 &
0.75 & 0.19 & 1800 \\ SL 2 & $00^{\rm h}17^{\rm m}44\fs0$ &
59\arcdeg00\arcmin33\arcsec & 0.25 & 0.08 & -332.1 & 1.3 & 8.3 & 2.5 &
2.19 & 0.63 & 0.13 & 5040\\ SL 3 & $00^{\rm h}17^{\rm m}44\fs8$ &
59\arcdeg00\arcmin15\arcsec & \multicolumn{2}{c}{$\cdots$} &
\multicolumn{2}{c}{$\cdots$} & \multicolumn{2}{c}{$\cdots$} & 0.32 &
0.88 & 0.22 & 1800 \\

\tablenotetext{a}{Errors are $1\sigma$, statistical. Systematic
$3\sigma$ calibration uncertainty is estimated to be $\sim\pm30$\%.}
\tablenotetext{b}{Values and errors derived from gaussian fit.}
\tablenotetext{c}{Integrated over a $\pm$FWHM range around the V$_{\rm
LSR}$ velocity. The CO ($J=3\rightarrow2$) V$_{\rm LSR}$ and FWHM were
used for the [C~I] non-detections.}  \tablenotetext{d}{Baseline
RMS. Computed using 3 MHz spectral resolution and scaled to T$_{mb}$.}
\tablenotetext{e}{Total accumulated integration time including on+off
positions.}  \enddata
\label{tabci}
\end{deluxetable}

\newpage

\begin{deluxetable}{lr@{$\pm$}lr@{$\pm$}lr@{$\pm$}lr@{$\pm$}lr@{$\pm$}lr@{$\pm$}lcrr@{$\pm$}l}
\tablewidth{0pc} \scriptsize \tablecaption{Measured $^{12}$CO line
parameters for IC~10-SE} \tablehead{ & \multicolumn{12}{c}{$^{12}$CO
($J=3\rightarrow2$)} & & \multicolumn{3}{c}{$^{12}$CO
($J=1\rightarrow0$)} \\ \cline{2-13} \cline{15-17} \\ &
\multicolumn{2}{c}{T$_{mb}$\ \tablenotemark{a,b}} &
\multicolumn{2}{c}{V$_{\rm LSR}$\ \tablenotemark{b}} &
\multicolumn{2}{c}{FWHM\ \tablenotemark{b}} &
\multicolumn{2}{c}{I$_{\rm CO}$\ \tablenotemark{a,f}} &
\multicolumn{2}{c}{I$_{\rm [C I]}$/I$_{\rm CO}$\ \tablenotemark{d,g}}
& \multicolumn{2}{c}{$\Delta v_{\rm CO}\over\Delta v_{\rm [C I]}$\
\tablenotemark{e}} & & \multicolumn{1}{c}{I$_{\rm CO}$\
\tablenotemark{c}} & \multicolumn{2}{c}{I$_{\rm [C I]}$/I$_{\rm CO}$\
\tablenotemark{d}}\\ & \multicolumn{2}{c}{(K)} &
\multicolumn{2}{c}{(km s$^{-1}$)} & \multicolumn{2}{c}{(km s$^{-1}$)}
& \multicolumn{2}{c}{(K km s$^{-1}$)}
&\multicolumn{2}{c}{}&\multicolumn{2}{c}{}& & (K km s$^{-1}$) \\ }
\startdata MC 1 & 1.66 & 0.02 & -329.7 & 0.1 & 13.2 & 0.2 & \ 22.8 &
0.8 & 0.13 & 0.02 & 1.9 & 0.3 & & 14.3\phn\phn\phn\phn & 0.20 & 0.04\\
MC 2 & 1.24 & 0.03 & -327.4 & 0.2 & 15.0 & 0.5 & 19.0 & 0.5 & 0.13 &
0.03 & 2.2 & 0.4 & & 11.7\phn\phn\phn\phn & 0.22 & 0.05\\ MC 3 & 0.77
& 0.03 & -336.5 & 0.4 & 18.5 & 0.9 & 14.9 & 0.6 & 0.15 & 0.04 & 1.9 &
0.5 & & 9.4\phn\phn\phn\phn & 0.24 & 0.06\\ SL 1 & 0.41 & 0.06 &
-332.6 & 0.6 & 8.1 & 1.3 & 3.4 & 0.5 & 0.11 & 0.22 &
\multicolumn{2}{c}{\ldots} & & 0.3\phn\phn\phn\phn &$<2.50$&2.70\\ SL
2 & 1.33 & 0.05 & -330.7 & 0.2 & 9.9 & 0.4 & 13.8 & 0.5 & 0.16 & 0.05
& 1.2 & 0.4 & & 5.5\phn\phn\phn\phn & 0.40 & 0.13\\ SL 3 & 0.88 & 0.04
& -327.9 & 0.4 & 17.0 & 0.8 & 15.5 & 0.6 & 0.02 & 0.05 &
\multicolumn{2}{c}{\ldots}& & 7.8\phn\phn\phn\phn &$<0.11$&0.12\\

\tablenotetext{a}{Errors are $1\sigma$, statistical.  Systematic
$3\sigma$ calibration uncertainty is estimated to be $\sim\pm30$\%.}
\tablenotetext{b}{Values and errors derived from gaussian fit.}
\tablenotetext{c}{OVRO integrated intensity convolved to JCMT [C~I]
beam size.  Its systematic $3\sigma$ calibration uncertainty is
estimated to be $\sim\pm30$\%.}  \tablenotetext{d}{The $1\sigma$ error
in the ratio includes the 10\% $1\sigma$ calibration uncertainty as
well as the statistical errors.}  \tablenotetext{e}{Ratio of
linewidths computed as ratio of the corresponding FWHM.}
\tablenotetext{f}{Integrated over a $\pm$FWHM range around the V$_{\rm
LSR}$ velocity.}  \tablenotetext{g}{Not corrected for beam size
difference.}  \enddata
\label{tabco}
\end{deluxetable}

\newpage

\begin{deluxetable}{lr@{$\pm$}lr@{$\pm$}lr@{$\pm$}lr@{$\pm$}lr@{$\pm$}lr}
\tablewidth{0pc} \scriptsize \tablecaption{Measured $^{13}$CO
($J=3\rightarrow2$) line parameters for IC~10-SE} \tablehead{ &
\multicolumn{2}{c}{T$_{mb}$\ \tablenotemark{a,b}} &
\multicolumn{2}{c}{V$_{\rm LSR}$\ \tablenotemark{b}} &
\multicolumn{2}{c}{FWHM\ \tablenotemark{b}} &
\multicolumn{2}{c}{I$_{\rm CO}$\ \tablenotemark{a}} &
\multicolumn{2}{c}{I$_{\rm^{12}CO}$/I$_{\rm^{13}CO}$} &
\multicolumn{1}{c}{RMS\ \tablenotemark{d}}\\ & \multicolumn{2}{c}{(K)}
& \multicolumn{2}{c}{(km s$^{-1}$)} & \multicolumn{2}{c}{(km
s$^{-1}$)} & \multicolumn{2}{c}{(K km s$^{-1}$)} &
\multicolumn{2}{c}{} & \multicolumn{1}{c}{(K)}\\ } \startdata MC 1 &
0.17 & 0.03 & -331.7 & 1.0 & 10.7 & 2.4 & 2.0 & 0.4 & 11.4 & 2.3 &
0.06 \\ MC 2 & 0.11 & 0.03 & -331.2 & 1.9 & 15.2 & 4.5 & 1.9 & 0.4 &
10.0 & 2.1 & 0.06 \\ MC 3 & 0.11 & 0.03 & -337.0 & 2.7 & 20.6 & 6.5 &
2.5 & 0.6 & 6.0 & 1.5 & 0.07 \\ SL 2 & \multicolumn{2}{c}{$\cdots$} &
\multicolumn{2}{c}{$\cdots$} & \multicolumn{2}{c}{$\cdots$} & 0.3 &
0.5\ $^{c}$ & \multicolumn{2}{c}{46\ \tablenotemark{e} \ \ \ \ } &
0.07 \\

\tablenotetext{a}{Errors are $1\sigma$, statistical.  Systematic
$3\sigma$ calibration uncertainty is estimated to be $\sim\pm30$\%.}
\tablenotetext{b}{Values and errors derived from gaussian fit.}
\tablenotetext{c}{Integrated over the velocity extent of the $^{12}$CO
line.}  \tablenotetext{d}{Baseline RMS. Computed using 3 MHz spectral
resolution and scaled to T$_{mb}$.}  \tablenotetext{e}{Formal errors
are $^{+\infty}_{-29}$.}  \enddata
\label{tab13co}
\end{deluxetable}

\newpage

\begin{deluxetable}{cr@{$\pm$}lr@{$\pm$}lcl}
\tablewidth{0pc} \scriptsize \tablecaption{Continuum measurements for
IC~10-SE} \tablehead{&\multicolumn{2}{c}{Wavelength\
\tablenotemark{a}}& \multicolumn{2}{c}{Flux\ \tablenotemark{a,b}} &
Telescope & \multicolumn{1}{c}{Reference} \\
&\multicolumn{2}{c}{($\mu$m)} & \multicolumn{2}{c}{(Jy)} &&}
\startdata &60 & 5 & 25.7 & 12.8 & IRAS HIRES\ \tablenotemark{c} &
This paper \\ &95 & 10 & 27 & 13.5 & KAO & Thronson et
al. 1990\protect\markcite{TH90} \\ &155 & 10 & 17 & 8.5 & KAO &
Thronson et al. 1990\protect\markcite{TH90} \\ &850 & 30 & 0.70 & 0.35
& JCMT & This paper \\ &1300 & 50 & 0.28 & 0.14 & IRAM & Wild
1998\protect\markcite{WI98} \\ \tablenotetext{a}{Errors are $3\sigma$
estimates.}  \tablenotetext{b}{Flux measured in a 50'' HPBW
beam. Errors correspond to adopted $\pm50$\% $3\sigma$ calibration
uncertainty.}  \tablenotetext{c}{Flux in a $60''\times50''$ HPBW beam
with P.A.=$38^\circ$.  HIRES resolution after 60 iterations is
$56''\times35''$ with P.A.=$38^\circ$.}  \enddata
\label{tabcont}
\end{deluxetable}

\newpage

\begin{deluxetable}{cr@{.}lr@{$\times$}llr@{$\times$}ll}
\tablewidth{0pc} \scriptsize \tablecaption{Continuum solutions for
IC~10-SE} \tablehead{&\multicolumn{2}{c}{Temperature}&
\multicolumn{2}{c}{$\tau_{\rm mm}$} & \multicolumn{1}{c}{$\beta$} &
\multicolumn{2}{c}{$N_{\rm H}\cdot b$} & \multicolumn{1}{c}{Comments}
\\ &\multicolumn{2}{c}{(K)}
&\multicolumn{2}{c}{}&&\multicolumn{2}{c}{(cm$^{-2}$)}&} \startdata &
58&6 & 5.2 & $10^{-5}$ & 0.31& 2.9&$10^{22}$ & general graybody
solution\\ & 54& $5^{+43.6}_{-17.1}$& $4.6^{+3.2}_{-2.4}$&$10^{-5}$ &
$0.48^{+0.46}_{-0.44}$ &$2.8^{+1.9}_{-1.4}$&$10^{22}$& graybody with
$S^{ff}_{\nu}=80\cdot\lambda_{\rm mm}^{0.15}$ mJy removed\
\tablenotemark{a}\\ & 38&3 & 1.4 & $10^{-5}$ & 1.5 & 8.1&$10^{21}$ &
two components model with fixed $\beta$\\ & 9&1 & 4.7 & $10^{-4}$ &
1.5 & 2.6&$10^{23}$ & \\ & 53&2 & 1.2 & $10^{-5}$ & 1.0 &
6.5&$10^{21}$ & two components model with fixed $\beta$\\ & 20&9 & 1.1
& $10^{-4}$ & 1.0 & 6.4&$10^{22}$ & \\ & 33&2 & 8.5 & $10^{-6}$ & 2.0
& 4.7&$10^{21}$ & two components model with fixed $\beta$\\ & 6&5 &
1.1 & $10^{-3}$ & 2.0 & 6.4&$10^{23}$ & \\ \tablenotetext{a}{Errors
are $3\sigma$ estimates based on a Monte Carlo analysis. Central
values are computed as the median of the Monte Carlo distribution.}
\enddata
\label{tabsol}
\end{deluxetable}

\end{document}